\documentstyle{article}
\begin{document}

\begin{center}
{\LARGE Quantum authentication protocol\\[0.5cm]}

{\small $^1$Guihua Zeng, and $^2$Guangcan Guo\\[0.1cm]
$^1$National Key Lab. on ISDN, XiDian University, Xi'an 710071, China\\
$^2$Department of physics and Nonlinear science centre, \\University of Science and 
technology of China, Heifei, 230026, China}
\end{center}

\begin{abstract}
In this letter, we proposed a quantum authentication 
protocol. The authentication 
process is implemented by the symmetric cryptographic scheme with quantum effects. \\
PACS:03.67.Dd,03.65.Bz\\[1.0cm]
\end{abstract}

\large

Since the first finding that quantum effects may protect privacy information transmitted 
in an open quantum channel by S.Wiesner [1], and then by C.H.Bennett and G.Brassard [2],
a remarkable surge of interest in the international scientific and industrial
community has propelled quantum cryptography into mainstream computer science
and physics [3-10]. Furthermore, quantum cryptography is becoming increasingly
practical at a fast pace [11,12]. 
Many quantum key distribution protocols have been proposed and some have been verified in experimental. However, the previous quantum key distribution (QKD) protocols are based 
on the legitimate users. In practice, the presented QKD protocols 
are completely insecure under the men-in-middle attack {\footnote{Men-in-middle attack:
When the legitimate communicator 
Alice communicates the legitimate communicator Bob, Eve 
intercepts all qubit sent by Alice, and communicates Bob with impersonating Alice. 
Finally, Eve obtains two keys $K_{AE}, K_{EB}$, where $K_{AE}$ represents the secret 
key between Alice and Eve, and $K_{AE}$ represents the secret key between Bob and Eve. 
As a result Eve can easily decrypt the ciphertext sent by Alice or Bob. }}.

In addition, the users' identities need to be verified in the point-to-point communicator and in the quantum cryptographic network [13], even if one does not need to generate the quantum key 
distribution. For axample, if the communicator Alice is a member of a network, when she want
to enter the network, she first need to input the her password or others characteristics to verify her identities. Alice can enter the network only her password is correct. This process 
is called the authentication. If the implementation is completed by the quantum effects,
we call it as quantum authentication.
Unfortunately, there is no known way to initiate quantum authentication.
Consider these cases, we propose a quantum authentication protocol in this letter,
which is implemented by using EPR pair with the Bell's theorem [14] (or the EPR 
correlation [15]). 

In classic cryptography, there are three classes protocols: the key distribution protocol, 
the key verification protocol, and the authentication protocol, the uses of these protocols 
are different. These conceptions are useful in quantum cryptography, correspondingly, they 
are quantum key distribution, quantum verification and quantum authentication. It needs to 
stress that the quantum authentication scheme is different from the quantum key verification scheme [16]. The quantum authentication scheme is always used in the case 
of the non-QKD like that in classic cryptography, e.g., when a user want to enter a 
network, he does not need to generate a key but need to input his correct identity or other
characteristics. However, the quantum key verification scheme is used for verification of the obtained key in the quantum key distribution.

In this  paper, We proposed a quantum authentication protocol. It executes the following steps:

Step 1. Alice and Bob transfer the sharing key $K_1$ {\footnote{Here we assume  Alice and Bob have a sharing key before the currently communication}} into a sequence of measurement basis.
While Alice and Bob need to verify their identification, or need to set up a new communication, 
they secretly transfer the reserved 
common key into a sequence of measurement basis according to the appointment. For example, 
if Alice and Bob use the measurement basis of polarization photon which was used in BB84 
protocol, they may let the bit '0' correspond to rectilinear measurement basis and '1' 
correspond to diagonal measurement basis, or vice versa.
We represent rectilinear measurement basis by the symbol $\oslash$, and represent diagonal 
measurement basis by the symbol $\odot$. After transferred, Alice and Bob obtain a sequence 
of measurement basis, respectively. For example, if  the common key is $K_1=001101$, the 
sequence of measurement basic is $M_{K_1}=\odot\odot\oslash\oslash\odot\oslash$.

Step 2. Alice and Bob set up a quantum communication channel. When Alice wants to secretly 
communicate Bob, Alice and Bob need to set up a quantum channel. The transmitting quantum 
states in the quantum channel may be arbitrary. For example the polarization photon state 
or the phase correction states[5]. In this protocol, we use the phase correction states. So the 
channel consists of a source that emits EPR pairs of spin-$\frac{1}{2}$ particle, in a singlet 
state. The particles fly apart along the $z$ axis, towards the two legitimate users of the 
channel. Alice chooses a random  basis for measuring one numbering of each EPR pair 
of particles. The other particle of each EPR pair is measured by Bob in the next step.
Alice's measurement results in effect determine, through  the EPR corrections, a sequence of 
states for Bob's particles.

Step 3. Bob measures the strings of quantum states.
Bob randomly measures the sequence of quantum states by using two 
measurement basis $M$ and $M_{K_1}$, where $M$ is the measurement basis for obtaining new authentication key and for quantum key 
distribution, $M$ is like the basis used in EPR protocol. 
$M_{K_1}$ is the measurement basis for identity authentication in the current communication. 

Step 4. Alice and Bob check the eavesdropper. 
For secure communication, the legitimate communicators Alice and Bob need to firstly detect 
the eavesdroppers. Bob randomly chooses some measurement results measured by the basis $M$ for 
checking the correction of EPR pair. Then the communicators judge the eavesdropping according 
to the Bell's theorem (or EPR correlation). 

Step 5. Bob encrypts his results measured by $M_{K_1}$.
Although Bob does not know the qubits measured by Alice, it will not influence the identity 
verification. Expressing the strings of quantum states for authentication by 
$$|\Psi>=\{|\psi_1>, |\psi_2>,\cdots, |\psi_n>\}.$$
Where $|\psi_i>$ represents a qubit received by Bob. After finished measurement, Bob obtains
$$|\Phi>=M_{K_1}|\Psi>,$$ 
where $|\Phi>=\{|\phi_1>, |\phi_2>,\cdots, |\phi_n>\}$ represents the measurement results 
under measurement basis $M_{K_1}$, i.e., $|\phi_i>=M_{K_1}|\psi_i>, i=1,2\cdots, n$. Transferring $|\Phi>$ into binary bits strings $m$, and 
then using $K_1$ to encrypt it, Bob obtains the ciphertext 
$$y=E_{K_1}(m).$$ 
Bob sends Alice the ciphertext $y$, and tells Alice the corresponding sequence numbers of
quantum states $|\psi_i>, i=1,2,\cdots, n$. 

Step 6. Verifying Bob's identity.
Having received Bob's results, Alice analyzes Bob's results. Alice decrypts the ciphertext, 
$$m=E^{-1}_{K_1}(y),$$ 
and compare her results with $m$, thereby Alice gets the measurement basis $M_{K_t}$. If
$K_t=K_1$, Bob's identity is true. 

Step 7 Verifying Alice' identity.
After Alice decrypted the ciphertext, Alice sends Bob the result $m'$. If $m'=m$,  
Alice's identity is true. 

Step 8. Alice and Bob discard the authentication keys $K_1$, and set up a new 
authentication keys. After finished authentication, the authentication key $K_1$ is no 
longer use. The legitimate users obtain a new authentication key $K_2$. 
The method is same as the quantum key distribution in the EPR protocol.

In practical communication, because of the noise effects errors are evitable in quantum 
channel. If the errors are produced by the Bob's measurement, Bob tell Alice the error
qubits to overcome the noise effects. If the errors are produced in transmission, Alice 
and Bob estimate the bound of errors $e_t$, and consider it in the identity verification. 
While the error is more than $e_t$, the communicators refuses each other, otherwise, the
communicators are legitimate.

The proposed scheme need a pre-key $K_1$, it means that an initial phase is necessary. 
Communicators may use quantum method to acquire the authentication key, e.g., the Biham's 
technology [13]. In the Ref.[13], Biham {\sl et al.} proposed a method to distribute the quantum key between Alice and Bob by the center. To prevent the center's cheating (men-in-middle attack), the center must be legitimate and believable.
Communicators can also use classic cryptographic method with quantum key distribution protocol 
to get the authentication key, e.g., the famous RSA system with QKD protocol. The method is as the following: first the legitimate communicators use the RSA system distributes the key, 
then in the valid time of RSA system obtain the sharing key $K_1$ by the QKD protocol
{\footnote{For practical quantum cryptography, the combination of quantum cryptography and classic cryptography perhaps be useful and pontential direction}}, 
here the `valid time' is important. 

The proposed quantum authentication protocol is provably secure. Because:
i) Our protocol does not have the conspiracy problem of masquerading. If a forger wants to 
masquerade user Alice or Bob to communicate with others, he must find the common key. However,
it is difficult to obtain the shared common secret because of the follows two reasons. 
First, the authentication key is obtained by the quantum key distribution protocol which 
is provably secure, so the authentication key is secure. Second the authentication key is 
used only one times, eavesdropper does not know any information about the authentication key.  
ii) The replay-attack will also not succeed in our protocol because the key is used only one 
times. iii) The quantum attacking strategy is invalid, the reason is the same as the analysis 
for previous QKD protocols.

There is a weakness in our protocol. Although the obtaining of the common
key in the last quantum communication is provably secure, the common key reservation has 
not circumvented possibility of attacking by eavesdroppers like in classic cryptography. 
In fact, this drawback exists in all symmetric cryptographic system.  Of 
course, we can use the EPR effects or other quantum effect, i.e., quantum memory, to keep 
the common key, but the reservation time is very short according to current 
technology. A long time correlation of quantum states is need in the future.

We use EPR effects with Bell' theorem (or EPR correlation) to implement quantum authentication. 
It can also be implemented by noncommute quantum states or non-orthogonal quantum states with Heisenberg uncertainty principle (including the mixed states) by using a similar procedure.

This project was supported by the Natural Science Foundation of China, Grant 
no: 69803008.

\eject
\begin{flushleft}
{\bf References}
\end{flushleft}

\begin{enumerate}
\item S. Wiesner, Sigact News, vol. 15, no. 1, 78 (1983). 
\item C. H. Bennett, G. Brassard, S. Breidbart, and S. Wiesner, Advances in Cryptology: 
Proceedings of Crypto 82, August 1982, Plenum Press, New York, pp. 267 - 275. 
\item C. H. Bennett, and G. Brassard, Advances in 
Cryptology: Proceedings of Crypto'84, August 1984, Springer - Verlag, pp. 475 - 480. 
\item C. H. Bennett, Phys. Rev. Lett. 68, 3121 (1992). 
\item A. K. Ekert, Phys. Rev. Lett. 67, 661 (1991). 
A. K. Ekert, J. G. Rarity, P. R. Tapster, and G. M. Palma, Phys. Rev. Lett. 69, 1293 (1992). 
\item C.H.Bennett,F.Bessette, G.Brassard, L.Salvail and J.Smolin,  J.Cryptology 5, 3 (1992). 
\item J. Breguet, A. Muller, and N. Gisin, J. Modern Optics, 41, 2405 (1994).
\item S. M. Barnett,  and S. J. D.Phoenix, Journal of Modern Optics, 40, 1443 (1993). 
\item C. A. Fuchs, N. Gisin, R. B. Griffiths, C. S. Niu, and A. Peres, Phys. Rev. A 56, 1163  (1997).
\item B. A. Slutsky, R. Rao, P. C. Sun, and Y. Fainman, Phys. Rev. A 57, 2383 (1998).
\item S. Phoenix, S. Barnett, P. Townsend and K. Blow, Journal of Modern Optics, 42, 1155 (1995).
\item C. Marand and P. D. Townsend, Optics Lett. 20, 1695 (1995).
\item E. Biham, B. Huttner, and T. Mor, Phys. Rev. A, Vol. 54, 2651 (1996).
\item J. S. Bell, Physics (Long Island City, N.Y.) 1, 195 (1965).
\item C. H. Bennett, G. Brassard, and N. D. Mermin, Phys. Rev. Lett. {\bf 68}, 557 (1992).
\item G. Zeng and W. Zhang, Phys. Rev. A, vol 61, no 1, 2000, (in press).
\end{enumerate}

\end{document}